\documentclass[aps,prb,showpacs,twocolumn]{revtex4}
\usepackage{graphicx}

\begin{document}
\title{Two-dimensional anisotropic Heisenberg antiferromagnet in a field}
\author{M.~Holtschneider}
\author{W.~Selke}
\author{R.~Leidl}
\affiliation{Institut f\"ur Theoretische Physik, Technische Hochschule,
             52056 Aachen, Germany}
 
\begin{abstract}
  The classical, square lattice, uniaxially anisotropic Heisenberg
  antiferromagnet in a magnetic field parallel to the easy axis
  is studied using Monte Carlo techniques.
  The model displays a long-range ordered antiferromagnetic, an algebraically
  ordered spin-flop, and a paramagnetic phase.
  The simulations indicate that a narrow disordered phase intervenes
  between the ordered phases down to quite low temperatures.
  Results are compared to previous, partially conflicting findings
  on related classical models as well as the quantum variant with spin S=1/2.
\end{abstract}

\pacs{75.10.Hk, 75.40.Mg, 75.40.Cx, 05.10.Ln}

\maketitle

%-----------------------------------------------------------------------------%
\section{Introduction}
The square lattice, uniaxially anisotropic Heisenberg antiferromagnet
in an external field $H$ parallel to the easy axis has been studied
since more than two decades both in its classical
version \cite{lb,gj,kst,cp} and in the quantum
variant.\cite{cu1,st} However, important features
of the phase diagram have not been definitely clarified.

The anisotropy may be introduced in various ways. In the XXZ case, the
Hamiltonian may be written as
\begin{equation}
  \label{Ham}
  \mathcal{H} = J \sum\limits_{(i,j)}\left[ \Delta (S_i^x S_j^x
   + S_i^y S_j^y) + S_i^z S_j^z\right]
   - H \sum\limits_{i} S_i^z,
\end{equation}
where for the classical model $S_i^x,S_i^y,S_i^z$ are the 
components of a unit vector corresponding to the spin at site $i$ 
of the lattice, and the sum $(i,j)$ runs over all nearest-neighbor 
pairs. The coupling constant $J$ and the field $H$ are positive; 
the anisotropy parameter $\Delta$ may range from zero to one. The 
model is known to exhibit, for $0 < \Delta < 1$, both for simple 
cubic and square lattices, an antiferromagnetic (AF) phase at low 
temperatures and low fields, a spin-flop (SF) phase with spins 
canted towards the $z$-axis at larger fields, and the disordered, 
paramagnetic phase.

For the simple cubic lattice, the three phases are believed to meet
at a bicritical point,\cite{fn} belonging to the universality
class of the isotropic Heisenberg model with O(3) symmetry. Elsewhere, the
boundaries to the disordered phase display either Ising criticality
in the AF case or XY criticality in the SF case.

For the square lattice, different scenarios on the fate of
the bicritical point have been discussed following the early
analysis of Landau and Binder.\cite{lb} In particular, (a) the
bicritical point may move to zero temperature, in accordance
with the well-known theorem by Mermin and Wagner,\cite{mw} with
a presumably narrow disordered phase in between the AF and SF
phases; (b) at low temperatures, there may be a direct transition
of first order between the AF and SF phases, with a tricritical
point on the boundary of the AF to the paramagnetic phase; and
(c) a 'biconical phase', in which the spins on only one of the
sublattices of the antiferromagnet are canted, may intervene between
the AF and SF phases at low temperatures. However, based on the
early simulations \cite{lb} at $\Delta=4/5$, none of these
three scenarios has been strongly or even definitely favored.

More recently, related two-dimensional antiferromagnets
in a field have been studied. For a classical antiferromagnet with
nearest-neighbor interactions and a single-ion anisotropy, it has
been suggested that the bicritical
point occurs at zero temperature. However, evidence is provided
by analytic approximations which are in quantitatively rather
poor agreement with Monte Carlo data.\cite{cp} A phase diagram
with a tricritical point and a direct transition between the
AF and SF phases has been determined for the quantum
variant of the XXZ-model with spin $S=1/2$ and $\Delta=2/3$, see
Eq.\ (\ref{Ham}), corresponding to the hard-core boson Hubbard
model.\cite{st} For an experimentally motivated classical
model \cite{mat} with single-ion anisotropy and further neighbor
couplings a topologically similar phase diagram has been obtained.\cite{ls}

Experimentally, there seem to be quite a few quasi two-dimensional
antiferromagnets with uniaxial anisotropy, including, for
instance, Rb$_2$MnF$_4$, Rb$_2$MnCl$_4$, K$_2$MnF$_4$,
La$_5$Ca$_9$Cu$_{24}$O$_{41}$, and
Mn(HCOO)$_2$.\cite{mat,gj,exp1,kst,exp2,exp3,exp4} However, the above
sketched subtleties of the phase diagram have not been 
fully elucidated, perhaps due to additional interactions like interlayer
couplings, or additional anisotropies like the breaking of
the XY isotropy, which, even when being weak, are expected to
affect significantly critical properties.\cite{cp} Inevitable
defects may eventually play an important role in two-dimensional
antiferromagnets in a field being related to random-field
systems.\cite{nat}

In this article, we shall present results of extensive Monte Carlo
simulations on the classical XXZ Heisenberg antiferromagnet
on a square lattice, mostly setting 
the anisotropy parameter to $\Delta=4/5$ and $\Delta=2/3$. Our findings
indicate that a narrow disordered phase intervenes between the AF
and SF phases down to quite low temperatures.

The layout of the article is as follows. In the next section, basic
properties of the model will be discussed and
the quantities computed in the simulations will be listed. The
transition from the SF to the paramagnetic phase will
be discussed in Sec.\ \ref{sec_sf}, followed by a section dealing with
the boundary line of the AF phase. A brief discussion and
summary will conclude the paper. 

%-----------------------------------------------------------------------------%
\section{Basic properties and quantities of interest}\label{sec_bp}
At zero temperature, $T= 0$, the XXZ model on a square lattice, see
Hamiltonian (1), may be easily solved exactly.\cite{lb} The
antiferromagnetic structure, being the ground state at low
fields, becomes unstable against the spin-flop state at
\begin{equation}
  \label{Hc1}
  H_{c1}= 4J \sqrt{1 - \Delta^2}.
\end{equation}
Increasing the field, the paramagnetic state gets stable at 
$H \ge H_{c2}$, with
\begin{equation}
  \label{Hc2}
  H_{c2}= 4J (1 + \Delta).
\end{equation}
The tilt angle, $\Theta$,  between the $z$-axis and the direction of the spins
in the SF state, $H_{c1} \le H \le H_{c2}$, is given by 
\begin{equation}
  \label{Theta}
  \Theta= \cos^{-1}[H/4J (1 + \Delta)].
\end{equation}

Obviously, in the isotropic limit, $\Delta=1$, one gets $H_{c1}=0$
and $H_{c2}=8J$, with the Ising-like AF state being squeezed out,
the ground state at $H=0$ having the rotationally
invariant O(3) symmetry. In the Ising limit, $\Delta =0$, one
has $H_{c1}=H_{c2}=4J$, i.e.\ the SF structure is certainly not
a ground state.

At low temperatures, considering $0 < \Delta < 1$, the AF
and SF states give rise to ordered phases. In the AF case, one expects
long-range order with the $z$-component of the sublattice
magnetization as the order parameter. The transition to the
paramagnetic phase is believed to be continuous and of Ising type, at
least at small fields. In the SF phase, the rotational invariance
of the $x$- and $y$-components of the canted spins may lead to
algebraic order and a transition of Kosterlitz-Thouless
type to the disordered phase,\cite{lb,cp,kt} as it is known to hold, for
instance, for the two-dimensional XY model.

\begin{figure}[ht]
  \includegraphics{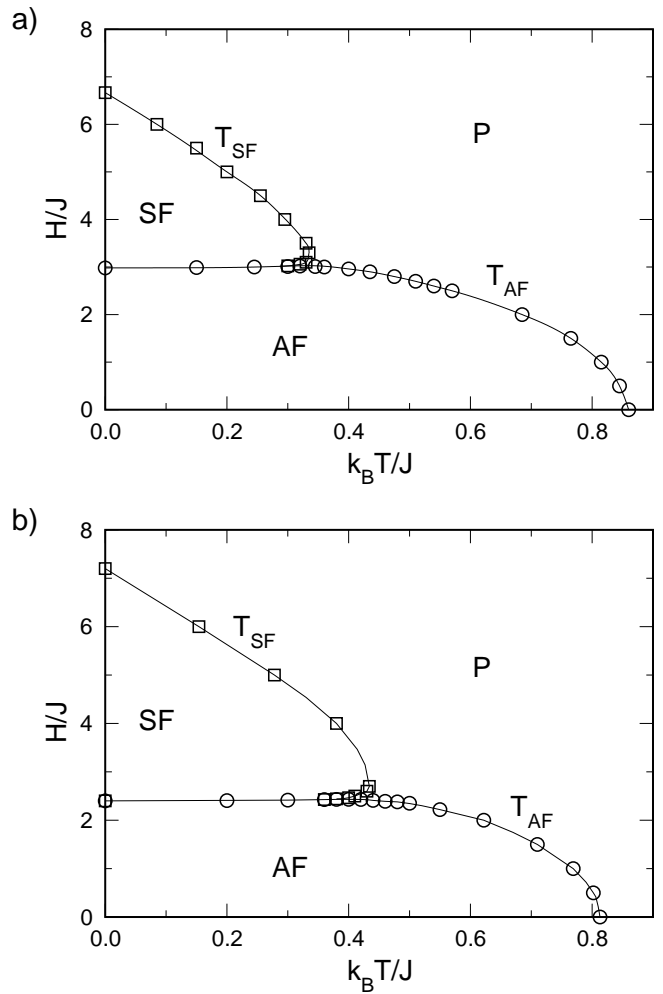}
  \caption{Simulated phase diagrams of the XXZ model on a square lattice,
   for the values (a) $\Delta=2/3$, and (b) $\Delta=4/5$
   of the anisotropy parameter. Squares denote transitions
   between the SF and paramagnetic phases, circles refer to the
   boundary of the AF phase. Lines, here and in the following
   figures, are guides to the eye.}
  \label{fig_phdiag}
\end{figure}

Phase diagrams of the XXZ model, as obtained from our Monte
Carlo simulations, are depicted in Figs.\ \ref{fig_phdiag}a
and \ref{fig_phdiag}b. We set $\Delta =4/5$ and $\Delta=2/3$,
to allow for comparison with previous findings.\cite{lb,st} The
diagrams will be discussed in detail later. Unless indicated
otherwise, here and in the following error bars
in the figures are smaller than the size of the symbols.

In the Monte Carlo (MC) simulations, we consider square lattices with 
$L \times L$ sites, employing full periodic boundary conditions. The
linear dimension $L$ ranges from 2 to 240, especially to
study finite-size effects allowing for extrapolation to
the thermodynamic limit, $L\longrightarrow\infty$ (see below). Applying
the standard Metropolis algorithm, each MC run consists
of at least $10^6$ (and up to $10^8$) Monte Carlo steps per site. To
obtain averages and error bars, we usually take into account
several, typically about ten (and up to forty), realizations choosing various
random numbers. In selected cases, especially at low
temperatures close to the boundaries of the AF and SF phases, the
MC runs are started  with different initial configurations to check for
correct equilibration.

We compute quantities of direct experimental interest as well
as other quantities which enable us to conveniently determine
the phase transition lines and critical properties. In particular, we
calculated the specific heat $C$, both from the energy fluctuations
and from the temperature derivative of the energy. Various
magnetizations were computed: Especially, we recorded (i) the
$z$-component of the total magnetization,
\begin{equation}
  <M^z> = \Big<\sum\limits_{i} S_i^z\Big> \big/ L^2;
\end{equation}
(ii) the square of the $z$-component of the staggered
magnetization \mbox{$<(M_s^z)^2>$} (or, similarly, the absolute value) 
to describe the order in the antiferromagnetic phase,
\begin{equation}
  \label{Mzst2}
  <(M_s^z)^2> =
   \Big< \Big[\sum\limits_{i_a}S_{i_a}^z
    - \sum\limits_{i_b}S_{i_b}^z \Big]^2\Big> \big/ L^2,
\end{equation}
summing over all sites, $i_a$ and $i_b$, of the two sublattices
$a$ and $b$ of the antiferromagnet; as well as (iii) the square of
the staggered transverse sublattice magnetization
\mbox{$<(M_s^{xy})^2>$}, to describe the ordering in the SF phase,
\begin{equation}
  \label{Mxy}
  <(M_s^{xy})^2> \; = \; <(M_s^x)^2> + <(M_s^y)^2>,
\end{equation}
in full analogy to Eq.\ (\ref{Mzst2}). Alternately, one may compute the
sum of the squares of each sublattice magnetization for both
transverse components, as before.\cite{ls}

In addition, we monitored the magnetic (staggered) susceptibility
$\chi_{(s)}^z$, which may be obtained from the fluctuations
of the (staggered) magnetization or its derivative
with respect to the (staggered) field.
To identify the type of transition along the boundary of
the AF phase, the fourth-order, size dependent cumulant $U_L^z$ 
of the staggered magnetization, the Binder cumulant,\cite{bin}
is supposed to be rather useful: 
\begin{equation}
  U_L^z = 1- <(M_s^z)^4>_L \big/ \left(3 <(M_s^z)^2>_L^2\right),
\end{equation}
where \mbox{$<(M_s^z)^4>$} is defined in analogy to \mbox{$<(M_s^z)^2>$}. The
corresponding cumulant $U_L^{xy}$ may be useful to study the boundary
of the SF phase. To identify the type of the phase transition, we
also study the fourth-order cumulant of the energy \cite{chal}
\begin{equation}
  \label{BindCum}
  V_L = 1- <E^4>_L \big/ \left(3<E^2>_L^2\right),
\end{equation}
with $E$ being the energy per site.

Further relevant information on
the phase transitions may be inferred from histograms of the
order parameter and of the energy.

%-----------------------------------------------------------------------------%
\section{Transition from the spin-flop to the paramagnetic phase}\label{sec_sf}
For square lattices, the spin-flop phase has been argued to be of
Ko\-ster\-litz-Thouless type,\cite{lb,cp,st} where transverse
spin correlations, i.e.\ $<S_{i}^xS_{i'}^x + S_{i}^yS_{i'}^y>$,
decay algebraically with distance $|i-i'|$ for widely separated sites 
$i$ and $i'$. Closely related, the transverse sublattice
magnetization \mbox{$<(M_s^{xy})^2>$}, see Eq.\ (\ref{Mxy}), describing the
ordering in the SF phase, is expected to behave
for $T>0$ and sufficiently large systems as
\begin{equation}
  \label{g}
  <(M_s^{xy})^2>_L\;\propto L^{-g},
\end{equation}
with $g$ approaching 1/4 at the transition from the SF
to the para\-magnetic phase,\cite{kt,pl} and $g=2$ in the para\-magnetic phase.
Thence, in two dimensions, the order parameter, \mbox{$<(M_s^{xy})^2>$},
vanishes in the SF phase
as $L \longrightarrow\infty$ at all temperatures $T>0$.

\begin{figure}[ht]
  \includegraphics{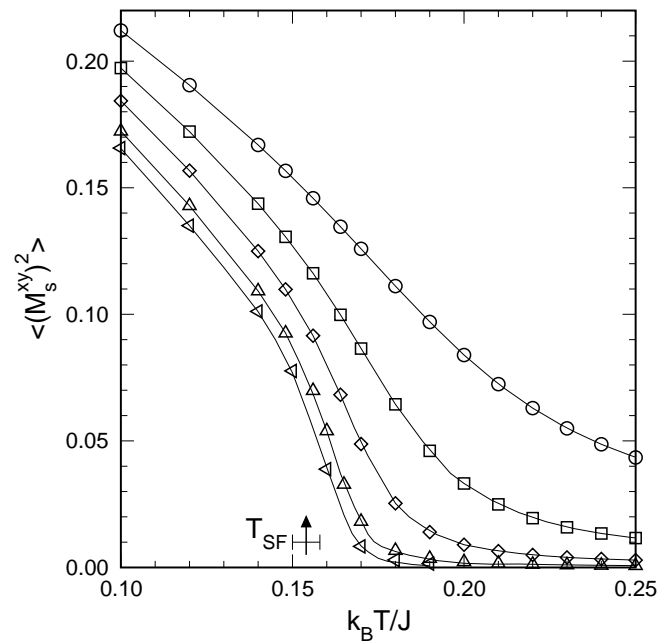}
  \caption{Square of the staggered transverse sublattice
   magnetization, \mbox{$<(M_s^{xy})^2>_L$}, vs.\ temperature at fixed field
   $H/J=6.0$ for systems of size $L = 10$, 20, 40, 80, and 120 (from top to 
   bottom), close to the  boundary between spin-flop and disordered
   phases. The anisotropy parameter is $\Delta=4/5$.}
  \label{fig_Mxy}
\end{figure}

In fact, as illustrated in Fig.\ \ref{fig_Mxy}, the
MC data clearly show that the magnetization decays
with system size both in the SF and the paramagnetic 
phase. To determine the boundary of the SF phase, $T_{\text{SF}}$, from
the size dependence of the transverse sublattice
magnetization, Eq. (10), one may
study the effective exponent, as usual,\cite{p}
\begin{equation}
  g_{\text{eff}}(L) = -\frac{\text{d}\ln<(M_s^{xy})^2>_L}{\text{d}\ln L},
\end{equation}
in its discretized form, comparing data for consecutive 
system sizes, $L_1$ and $L_2$, $L_2 > L_1$, with
\begin{equation}
  g_{\text{eff}}(L_0) =
   - \frac{\ln[<(M_s^{xy})^2>_{L_2}/<(M_s^{xy})^2>_{L_1}]}{\ln (L_2/L_1)},
\end{equation}
where $L_0= \sqrt{L_1 L_2}$. Indeed, when crossing the phase
boundary by, for instance, fixing the field
and decreasing the temperature (see Figs. 1a and 1b), for large
systems $g_{\text{eff}}$ 
tends to drop rapidly from rather high values approaching 2, characterizing
the decay in the disordered phase, to a small value close
to 1/4 at the transition to the SF phase. Deeper in the SF
phase, $g_{\text{eff}}$ decreases somewhat to even lower values.

\begin{figure}[ht]
  \includegraphics{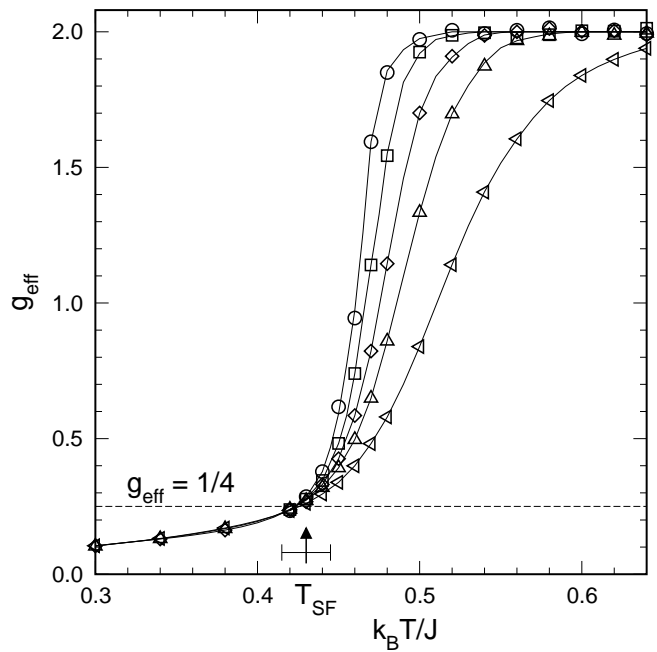}
  \caption{Effective exponent $g_{\text{eff}}$, see Eqs. (11)
   and (12), vs.\ temperature at fixed field
   $H/J=2.7$, comparing systems of
   size $(L_2,L_1)$= (20,10), (30,20), (40,30), (60,40), and
   (80,60) (from bottom to top), close to the
   boundary between spin-flop and disordered phases. The anisotropy
   parameter is $\Delta=4/5$.}
  \label{fig_geff}
\end{figure}

This behavior of the effective exponent is exemplified in
Fig.\ \ref{fig_geff}, allowing one to estimate $T_{\text{SF}}$. We checked
the approach by analyzing the two-dimensional (planar)
XY model, reproducing quite accurately the transition temperature
as obtained from elaborate numerical studies.\cite{b}

The Kosterlitz-Thouless character of the transition between the SF
and the paramagnetic phases is also reflected in the thermal behavior
of the specific heat $C$, which displays a non-critical maximum close to,
but not exactly at the transition. Of course, from simulational data
one cannot identify the expected essential singularity of $C$
at the transition.

Approximate analytic expressions for the boundary between the SF
and the paramagnetic phases may be obtained by the following
considerations, slightly modifying previous
arguments.\cite{lb} Using polar coordinates
and fixing the $z$-component of the spins in the SF phase to its
value in the ground state, Eq.\ (\ref{Theta}), the transition
temperature $T_{\text{SF}}$ may be approximated by \cite{lb}
\begin{equation}
  \label{Tsf_approx}
  \frac{k_BT_{\text{SF}}}{J} = \frac{k_BT_{\text{KT}}}{J}
    \,\Delta\left[1-(H/H_{c2})^2\right],
\end{equation}
where $k_BT_{\text{KT}}/J$ refers to the Kosterlitz-Thouless transition
temperature of the XY model with coupling $J$, i.e.\ the classical isotropic
model with a two-component spin vector of length one.\cite{b} Indeed, this
approximate expression may be regarded as an upper bound to the
true transition temperature, because the fluctuations
in the $z$-component (which, in turn, are coupled to the
fluctuations in the transverse spin components) tend to reduce the 
transition temperature. In addition, the
change of the tilt angle $\Theta$ with temperature, at fixed field, will
affect the transition temperature. In fact, $T_{\text{SF}}$, as determined
in the simulations, see Figs.\ \ref{fig_phdiag}a and \ref{fig_phdiag}b,
is seen to be lower than suggested by Eq.\ (\ref{Tsf_approx}). In any event,
the approximation does not hold in the vicinity of $H_{c1}$,
where the fluctuations of the longitudinal and transverse  spin components
are strongly correlated. Among others, that region will be discussed
in the next section.

%-----------------------------------------------------------------------------%
\section{The boundary line of the antiferromagnetic phase}\label{sec_af}
We determined the boundary line of the AF phase
by monitoring especially the specific heat $C$,
the square of the $z$-component of the
staggered magnetization, \mbox{$<(M_s^z)^2>$}, the
staggered susceptibility, $\chi_s^z$, and the Binder
cumulant, $U_L^z$. The transition temperatures follow from finite-size
extrapolations of the Monte Carlo data. Results for
$\Delta=4/5$ and $\Delta=2/3$ are displayed in Fig.\ \ref{fig_phdiag}.

At low fields and high temperatures the transition is expected
to be continuous and of Ising type. Indeed, we observe, for
instance, a logarithmic divergence in the height of the
peak in the specific heat as a function of system
size, $C_{\text{max}} \propto \ln L$, and an effective
critical exponent of the order parameter $\beta_{\text{eff}}$ consistent
with the asymptotic value 1/8 when approaching the transition, fixing
the field and varying temperature.

\begin{figure}[ht]
  \includegraphics{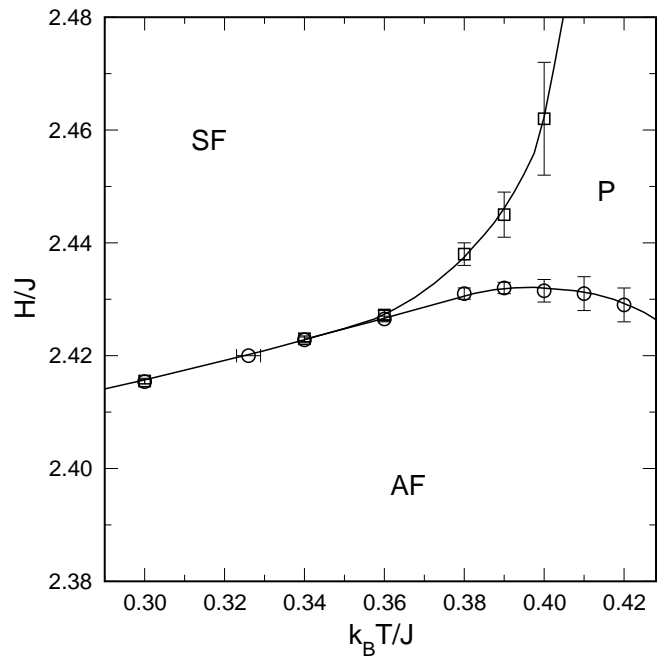}
  \caption{Details of the phase diagram of the XXZ model on a square
   lattice with $\Delta=4/5$. Squares refer to the boundary of the SF,
   circles to that of the AF phase.}
  \label{fig_phdiag_detail}
\end{figure}

The transition line of the AF phase shows a maximum
in the critical field, as a function of temperature, being somewhat
higher than $H_{c1}$, Eq.\ (\ref{Hc1}), both for $\Delta=2/3$
and $\Delta=4/5$ (a feature which has not been mentioned in the
early work\cite{lb}). In that part of the phase diagram,
see Fig.\ \ref{fig_phdiag_detail} for $\Delta=4/5$, the boundary lines
of the SF and AF phases approach each other, when 
lowering the temperature.

\begin{figure}[ht]
  \includegraphics{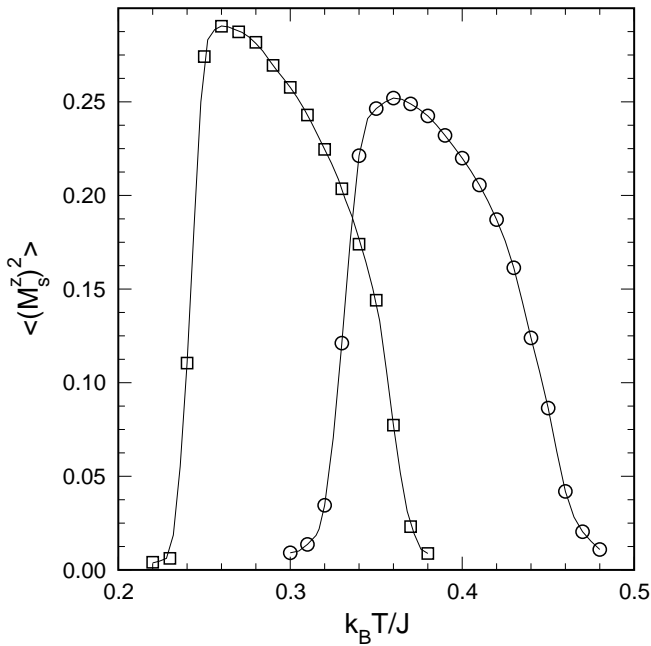}
  \caption{Staggered magnetization \mbox{$<(M_s^z)^2>$}
   vs.\ temperature at field
   $H/J=3.0$ for $\Delta=2/3$ with $L=128$ (squares), and
   at $H/J$= 2.42 for $\Delta=4/5$ with $L= 120$ (circles), close
   to the maximum of the boundary of the AF phase.}
  \label{fig_Mzst}
\end{figure}

Fixing the field to be slightly below its maximal
critical value and decreasing the temperature, one
observes a rather asymmetric behavior in the staggered
magnetization \mbox{$<(M_s^z)^2>$}, as illustrated in Fig.\ \ref{fig_Mzst}.
The order parameter of the AF phase rises fairly gradually entering
the AF phase from the paramagnetic phase, while
it drops down rather rapidly on approach to the SF phase. This
asymmetry signals either a change in the type of the
transition at the lower temperature, becoming, possibly, of
first order, or in the extent of the critical region
becoming, possibly, very narrow.

\begin{figure}[ht]
  \includegraphics{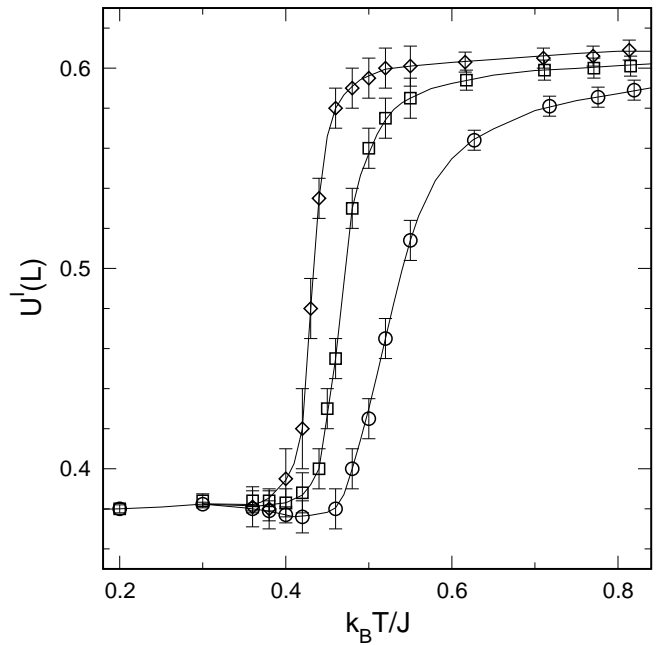}
  \caption{Binder cumulant $U^I(L)$, as obtained from the
    intersection points of cumulants for lattices of sizes $L$
    and $2L$ close to the boundary of the AF phase, as a function
    of temperature, with $L$= 10, 20, and 40 (from bottom to
    top). The anisotropy parameter is $\Delta=4/5$.}
  \label{fig_BindCum}
\end{figure}

To study the transition along the boundary of the AF phase in more detail,
the Binder cumulant $U_L^z$ may be quite useful, as had been employed
before.\cite{st,ls} In the thermodynamic limit the value of the cumulant
at the transition point, the critical cumulant
$U_{L=\infty}^z(T_{\text{AF}})$, is believed to characterize
the type and universality class
of the transition. From simulational data, the critical cumulant
may be estimated from the intersection value
$U^I(L)$ of the cumulant for different system sizes
$L$ and $L'= bL$ (in the following, we set $b=2$), doing
finite-size extrapolations.\cite{bin} In Fig.\ \ref{fig_BindCum},
we depict results for $\Delta=4/5$, monitoring $U^I(L)$ as
a function of temperature in the vicinity of the AF 
boundary line.
The intersection value $U^I$ seems to display
one plateau at higher temperatures and another plateau, with a smaller
height, at lower temperatures, with a fairly rapid change in between. The
plateaus tend to get more pronounced and the change becomes sharper
as the system size $L$ is increased. In fact, the plateau at
high temperatures is obviously related to the critical
cumulant in the universality class of the two-dimensional
Ising model, where
$U_{L=\infty}(T_{\text{AF}}) \approx 0.6106$.\cite{bn} Note 
that $U_L^z$, computed exactly at the
transition $(T_{\text{AF}},H_{\text{AF}})$, approaches this value closely with
only minor finite-size corrections for low fields and high
temperatures. However, near the maximum of the
boundary line of the AF phase, significant corrections are
observed, which have not been discussed in the related
quantum variant.\cite{st} The critical cumulant of the apparent
plateau at low temperatures
tends to increase weakly with T, being close to 0.4 at $\Delta=4/5$,
for the system sizes we considered, $L \le 120$. Perhaps interestingly, at
the critical field $H_{c1}(\Delta)$, the cumulant seems to approach,
at least for small lattices, in  the limit $T\longrightarrow 0$, a value close
to 0.38. With decreasing $\Delta$, this limiting value decreases.

\begin{figure}[ht]
  \includegraphics{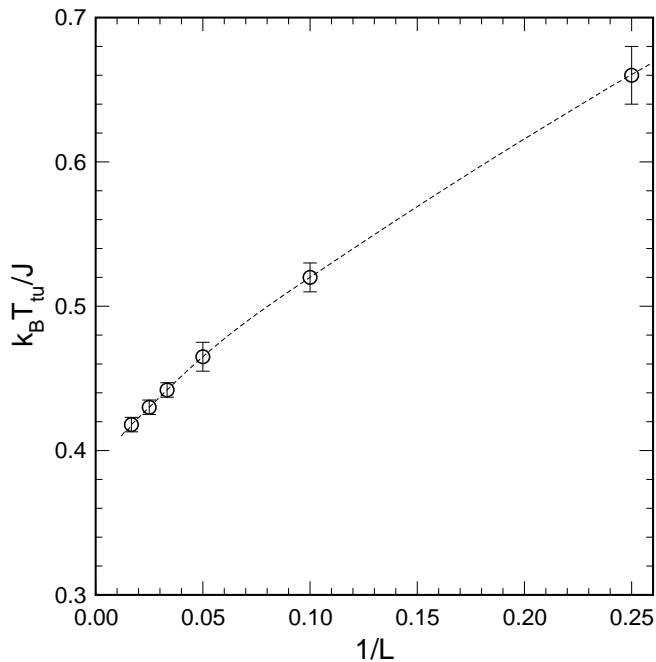}
  \caption{Size dependence of the turning point, $T_{\text{tu}}$, in the
    Binder cumulant $U^I(L)$ comparing lattices of linear
    dimension $L$ and $2L$, for $\Delta=4/5$, see Fig.\ \ref{fig_BindCum},
    with $L$ ranging from 4 to 60.}
  \label{fig_tu}
\end{figure}

Note that the turning point of the intersection value $U^I(L)$
is shifted towards lower temperatures, $T_{\text{tu}}$, as $L$
increases, see Figs.\ \ref{fig_BindCum} and \ref{fig_tu}. Obviously,
there are uncertainties in extrapolating those data to the thermodynamic
limit. A naive linear extrapolation yields an estimate of
a temperature, $T^*$, of about 0.4 (in units of $J/k_B$), at which
non-Ising criticality may set in, possibly due to a transition
of first order.

This interpretation, however, has to be viewed with
much care. Indeed, the fourth order energy cumulant \cite{chal} gives
no hint of a transition of first order at the AF
boundary, at quite low temperatures, $0.3J/k_B \leq T <T^*$. Its
minimum near the transition, becoming even more
shallow at lower temperatures, tends to vanish for larger lattices, as
expected for a continuous transition.

\begin{figure}[ht]
  \includegraphics{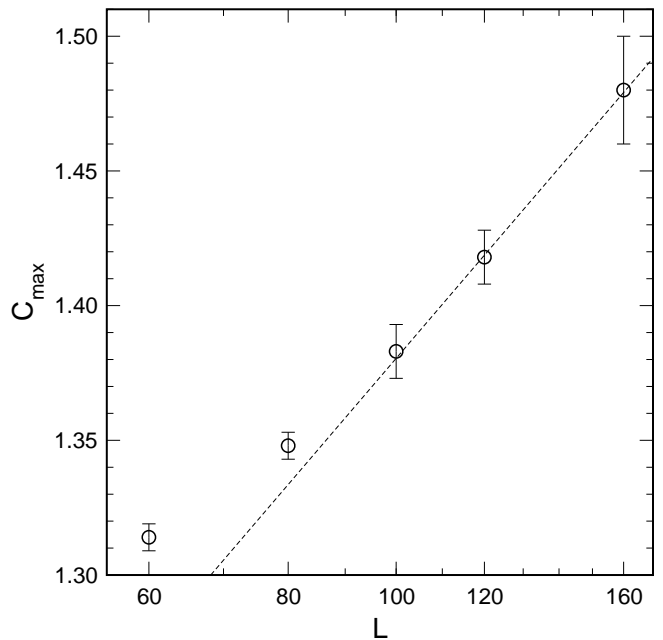}
  \caption{Maximum in the specific heat vs.\ logarithm of the
    system size $L$ at $H/J= 2.42$ and $\Delta= 4/5$ at 
    temperatures close to $k_BT_{\text{AF}}/J \approx 0.33$, see  
    Fig. 4. The dashed line indicates a logarithmic size dependence.}
  \label{fig_spheat}
\end{figure}

\begin{figure}[ht]
  \includegraphics{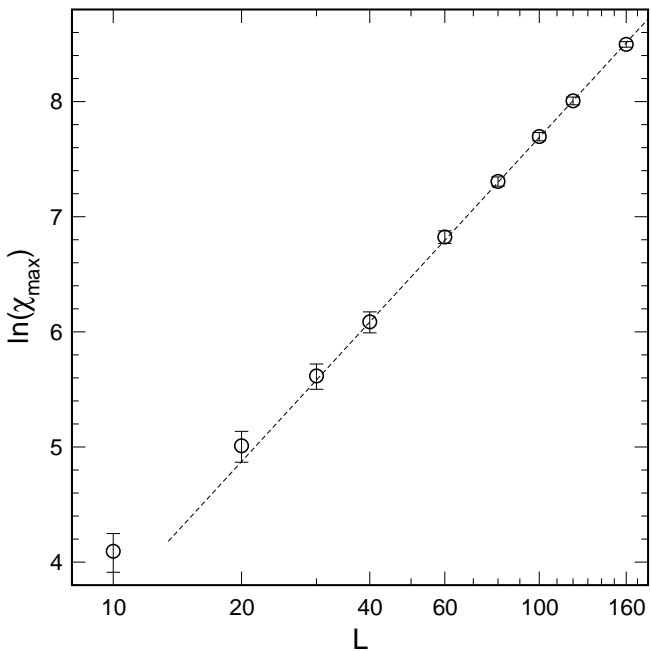}
  \caption{Doubly logarithmic plot of the size dependence of the
    maximum in the staggered susceptibility
    at $H/J= 2.42$ and $\Delta= 4/5$. The dashed line refers to a
    slope with $\gamma /\nu= 7/4$.}
  \label{fig_staggsusc}
\end{figure}

Even more strikingly, the size dependences of the
specific heat as well as of the staggered susceptibility suggest
that the boundary line of the AF phase is still in the Ising
universality class well below the anticipated change at $T^*$.
This behavior is exemplified in Figs.\ \ref{fig_spheat} and \ref{fig_staggsusc}
for the case $\Delta=4/5$ and $H/J=2.42$,
where $k_BT_{\text{AF}}/J \approx 0.33$, i.e.\ in the part of the phase diagram
where the AF and SF phase boundaries are hardly discernible,
see Fig.\ \ref{fig_phdiag_detail}.
We find consistency with  Ising criticality, especially,
$C_{\text{max}} \propto \ln L$, and the peak of the staggered
susceptibility, $(\chi_s^z)_{\text{max}}$, grows
like $L^{\gamma/\nu}= L^{7/4}$. 
We checked equilibration of the MC data by
monitoring time series of the energy and the staggered magnetization
as well as by calculating the specific heat from the
energy fluctuations and the temperature derivative of
the energy. The findings suggest that there is
a very narrow paramagnetic phase intervening between
the AF and SF phases at least down to that
temperature, $k_BT/J \approx 0.33$. At even lower
temperatures, this type of analysis would require very
large computational resources to obtain reliable MC data.

For $\Delta=2/3$, similar features are observed in the corresponding part
of the phase diagram, where, e.g., a logarithmic size dependence of
the height of the specific heat at temperatures close to
$k_BT_{\text{AF}}/J \approx 0.28$ is found, compare Fig.\ \ref{fig_phdiag}.

In contrast, results of our simulations of the XXZ model
on a \textit{simple cubic} lattice, setting $\Delta=4/5$, show
at temperatures below the bicritical point
a direct transition of first order between the AF and SF
phases, as expected.\cite{fn,bl2} For instance, the maximum
of the specific heat increases with a power law in
$L$, with an effective exponent, for moderate system
sizes, $8 \leq L \leq 20$, being significantly larger
than the value characteristic
for the universality class of the three-dimensional
Ising model. The behavior is, indeed, indicative of a
first order transition, where $C_{\text{max}} \propto L^3$. Likewise, 
energy histograms show the usual properties of a transition of first
order, with an overlap of two distinct Gaussian peaks. The critical
energy cumulant in the three-dimensional
case is observed to be close to, but distinct from 2/3.

Note that at fixed low temperature and varying the field, the
$z$-component of the magnetization changes rapidly in both sublattices
simultaneously, within our field resolution, both for $\Delta=4/5$
and $\Delta=2/3$, contradicting a biconical intermediate phase with a canting
of the spins in only one sublattice.

%-----------------------------------------------------------------------------%
\section{Discussion and summary}
We have studied the classical, square lattice, uniaxially anisotropic
Heisenberg antiferromagnet, the XXZ model, in a magnetic field, doing 
extensive Monte Carlo simulations. We mainly considered two
cases of fairly weak anisotropy, $\Delta=4/5$ and $\Delta=2/3$.

The model displays an antiferromagnetic, a
spin-flop, and a paramagnetic phase.

The transition from the antiferromagnetic to the paramagnetic phase
is seen to be of Ising type at small fields, remaining of that
type at higher fields and rather low temperatures (down to at
least $k_BT/J \approx 0.33$ in the case $\Delta= 4/5$), as
inferred, especially, from the size dependence of the maxima in the
staggered susceptibility and the specific heat. A naive analysis of MC 
data of the Binder cumulant along the boundary
of the antiferromagnetic phase may lead to a different, but
supposedly erroneous
conclusion. Presumably, the critical region at the AF phase
boundary becomes rather tiny when the SF and AF phase boundaries
approach each other, demanding very large lattices to
identify the universality class correctly, at least when
studying the Binder cumulant.

The transition between the spin-flop and the disordered phase
is of Kosterlitz-Thouless type. It may be located accurately
by analyzing the finite-size behavior of the transverse staggered
magnetization, vanishing with increasing system size with a
characteristic power law at the transition.

In the quantum variant of the model with
spin $S= 1/2$ a tricritical point on the
AF boundary and a direct transition between the AF and SF phases at
low temperatures have been obtained.\cite{st} Near the tricritical point the 
AF and SF phases are still well separated. At present, we
can only speculate whether this, perhaps, surprising
discrepancy between the classical and quantum models
could be resolved when the classical version would be
analyzed at even lower temperatures or when MC data on
the quantum version would be reanalyzed.

In  the isotropic limit, $\Delta=1$, one finds, in
non-vanishing field, $H > 0$, a spin-flop phase of Kosterlitz-Thouless
character both in the classical and in the
quantum model.\cite{lb,cu2} For $\Delta < 1$ and $H= 0$, an Ising type
transition occurs at a non-zero temperature, with
the transition temperature moving to zero as the anisotropy
vanishes, $\Delta \longrightarrow 1$, again both in the classical
and the quantum case.\cite{bar,cu1}  

One of the remaining open crucial questions is, whether there is
a typical phase diagram for square lattice, uniaxially
anisotropic antiferromagnets. In the case of the
antiferromagnetic nearest-neighbor Heisenberg model with
single-ion anisotropy, a phase diagram with
a paramagnetic phase between the antiferromagnetic and
spin-flop phases extending down to
arbitrarily low temperatures has been
suggested.\cite{cp} In another study of that model, a conflicting
scenario with a direct transition between the AF and SF phases
has been favored, presenting, however, only few Monte
Carlo data.\cite{kst} Including
more than nearest-neighbor interactions in an antiferromagnet
with single-ion anisotropy,\cite{mat,ls} or taking
into account quantum fluctuations,\cite{st} phase
diagrams with a tricritical point and a direct transition between
the AF and SF phases have been obtained.

In any event, we should like to encourage future work on
this and related models, which may also serve as a guide in
interpreting experiments on corresponding
quasi two-dimensional anisotropic antiferromagnets.

\acknowledgments
It is a pleasure to thank M.\ Troyer for a useful
conversation. Financial support by the Deutsche Forschungsgemeinschaft
under grant No.\ SE324 is gratefully acknowledged.

\end{document}